\documentclass[aip,rsi,preprint,graphicx]{revtex4-1} 
\usepackage{graphicx}
\usepackage{dcolumn}
\usepackage{bm}
\usepackage{hyperref}

\begin{document}

\title{A broadband microwave Corbino spectrometer at $^3$He temperatures and high magnetic fields}

\author{Wei Liu}
\affiliation{Department of Physics and Astronomy, The Johns
Hopkins University, Baltimore, MD 21218}
\email{liuwei@pha.jhu.edu}
\author{LiDong Pan}
\affiliation{Department of Physics and Astronomy, The Johns
Hopkins University, Baltimore, MD 21218}

\author{N.P. Armitage}
\affiliation{Department of Physics and Astronomy, The Johns
Hopkins University, Baltimore, MD 21218}

\date{\today}

\begin{abstract}
We present the technical details of a broadband microwave spectrometer for measuring the complex conductance of thin films covering the range from 50 MHz up to 16 GHz in the  temperature range 300 mK to 6 K and at applied magnetic fields up to 8 Tesla. We measure the complex reflection from a sample terminating a coaxial transmission line and calibrate the signals with three standards with known reflection coefficients. Thermal isolation of the heat load from the inner conductor is accomplished by including a section of NbTi superconducting cable (transition temperature around 8 $-$ 9 K) and hermetic seal glass bead adapters. This enables us to stabilize the base temperature of the sample stage at 300 mK. However, the inclusion of this superconducting cable  complicates the calibration procedure.  We document the effects of the superconducting cable on our calibration procedure and  the effects of applied magnetic fields and how we control the temperature with great repeatability for each measurement. We have successfully extracted reliable data in this frequency, temperature and field range for thin superconducting films and highly resistive graphene samples.
\end{abstract}


\maketitle
\section{Introduction}
Microwave  Corbino spectrometers are capable of providing complex broadband spectral information in the microwave regime \cite{booth94a}.  This  technique measures the complex optical response  as a function of frequency without resorting to Kramers-Kronig transforms. The method was used  by Anlage's group to  study  thin film high temperature superconductors \cite{wu95,anlage96,BoothPRL96a,xu09} as well as colossal magnetoresistive manganites \cite{schwartz00}.  Later, another group applied the same technique to study the microwave AC conductivity spectrum of a doped semiconductor \cite{LeePRL01b}. This technique has also been used to study the dielectric response of liquids and soft condensed matter \cite{martens00}. The Stuttgart group reported reliable measurements down to 1.7 K \cite{scheffler05b}. They measured the  frequency dependence of microwave conductivity of  heavy fermion metals \cite{Scheffler05a} and superconducting Al films \cite{Scheffler08a}. Kitano et al. constructed a Corbino spectrometer \cite{kitano08a} to investigate the critical behavior of LSCO \cite{kitano07a,KitanoPRB09a} as well as NbN films \cite{ohashi06}. Recently, a Corbino spectrometer that goes down to 2.3 K has been constructed to study NbN films \cite{mondal13}.

In this article, we describe the design of our microwave spectrometer to measure the complex conductance of thin films of InO$_x$ \cite{liu11a,liu13a} and graphene \cite{liu11b} at $^{3}$He temperatures and high magnetic fields. Related measurements at $^{3}$He  temperatures have been also reported by the Stuttgart  group \cite{steinberg12a}. In our group, the spectrometer covers the frequency range between 0.05 GHz and 15 GHz down to 300 mK and at  magnetic fields  up to 8 T. We demonstrate that we can repeat the measurements in a very reliable fashion over the interesting temperature and field ranges. As a non-resonant technique, the spectrometer requires an intricate and careful calibration procedure and presents a number of experimental challenges. In this paper, we  address  a number of  difficulties peculiar to this frequency range that must be overcome for measurements at these low temperatures and high magnetic fields.

\section{Experimental setup overview}
\begin{figure}
 \includegraphics[width=\columnwidth]{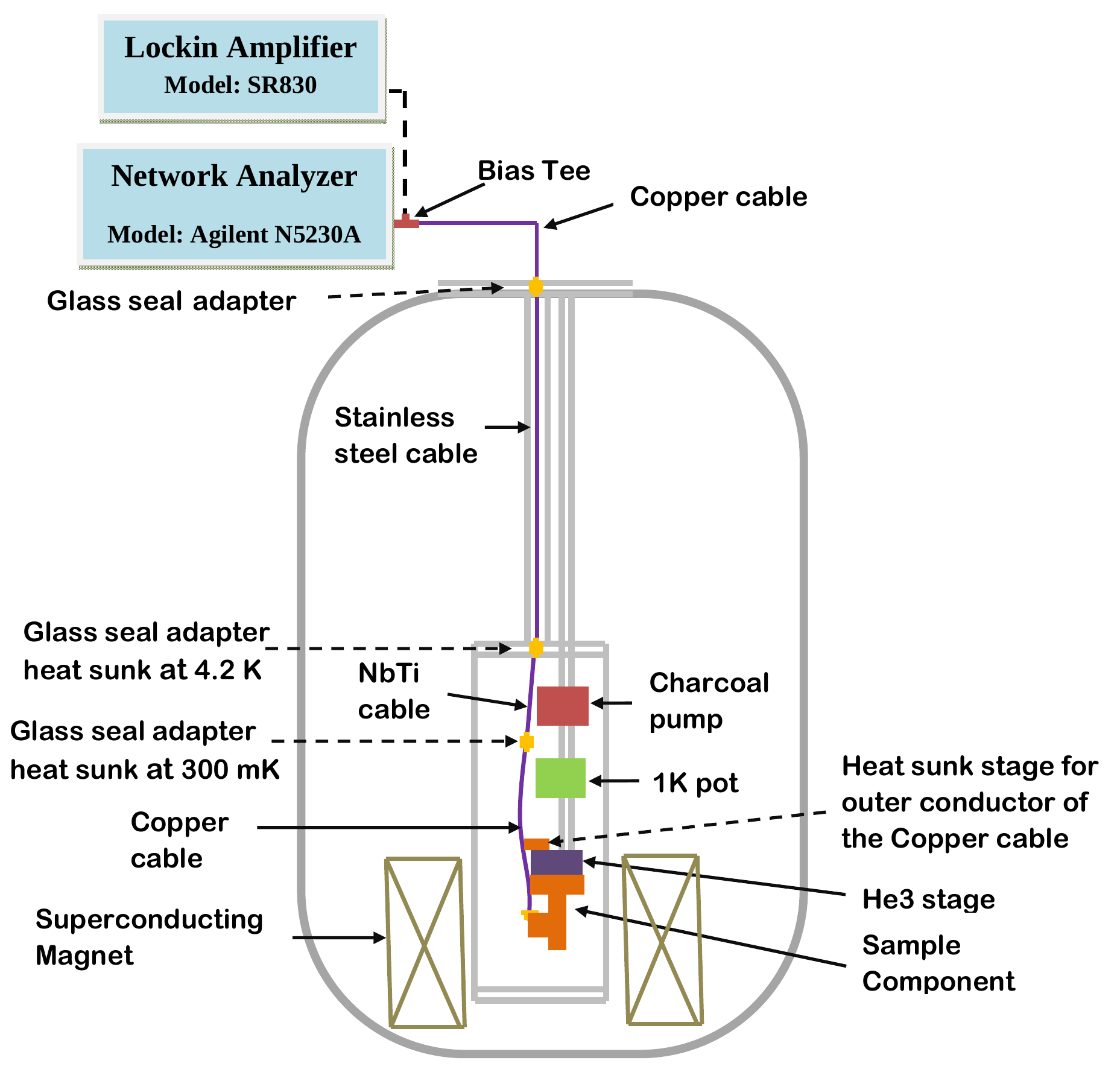}
 \caption{Schematic of the  experimental setup showing all the coaxial cables sections and connections in the Corbino microwave spectrometer.}
 \label{Setup}
 \end{figure}

A schematic drawing of the experimental setup is shown in Fig. \ref{Setup}. Microwave radiation is generated by a vector network analyzer \cite{Agilent N5230A}, and  guided by coaxial cables. The radiation comes down along  the coaxial cables and is reflected by the sample that terminates the transmission line. The reflected signal travels back along the same coaxial cable and is analyzed by the network analyzer. Losses and reflections in the transmission lines are calibrated  by a calibration procedure that is discussed in section IV.

Four sections of semi-rigid coaxial cables are used. The overall length of the whole transmission line is roughly about 1.4 meters. Copper coaxial cables \cite{Micro-Coax Company,  UT-85C-TP-LL} are used outside the cryostat connecting the network analyzer and also are used to connect the superconducting cable and the Corbino probe. Inside the cryostat, the upper longest section in the transmission line is the stainless steel coaxial cable \cite{Micro-Coax Company, UT-085-SS}.  A 10 cm long NbTi superconducting coaxial cable \cite{Keycom Company, NbTiNbTi085A} was added into the system to isolate the heat load from the room temperature connections, especially  the inner conductor.

A particular experimental challenge in performing these experiments in a $^3$He   environment  was the isolation of the heat load from the top of the coaxial line, especially the heat load from the inner conductor. We thermally anchor the outer conductor of the transmission line at two locations: the top flange of the inner vacuum chamber (IVC) at 4.2 K and the $^3$He pot at base temperature 300  mK. To ensure good thermal contact, a glass seal adapter (discussed below) is inserted in-line, attached to a copper housing and bolted securely against the top flange of the IVC. Another copper housing for a glass seal adapter between NbTi cable and copper cable was bolted securely to the top of the $^3$He pot.  This guarantees  that the outer conductor of  the superconducting cable is thermally anchored at 4.2 K and at the base temperature of the cryostat as shown in Fig. \ref{Setup}. Copper wires and  thermal grease around the cables, connectors and adapters  are also used for the connection between NbTi and copper cables to ensure that the outer conductors of the cables are well thermally anchored along the connection.

It is  more difficult  to heat sink the inner conductor due to the poor thermal conductivity of  teflon, which is in general the dielectric for the coaxial cables and adapters.  In initial measurements, before we added the superconducting coaxial cable, the base temperature of the setup was only 500 mK and the holding time was less than half an hour.  To make thermal contact to the center conductor, the three hermetically sealed glass bead adapters \cite{Kawashima Manufacturing Company, KPC185FFHA} were incorporated as displayed in Fig. \ref{Setup}. We have found that the glass bead inner dielectric in these adapters conduct heat much better than the teflon dielectric in conventional coaxial cables. As mentioned above, two of these special adapters were heat sunk respectively at 4.2 K and the $^3$He stage that were separated by the 10 cm long superconducting NbTi coaxial cable as showed in Fig. \ref{Setup}. The transition temperature of the superconducting cables is  8 $-$ 9 K. All the cables, glass bead adapters and connectors (except the section outside the dewar which stays at room temperature for all the measurements) were thermal cycled several times in liquid nitrogen before and after assembly to reduce the effects of thermal contractions  in later measurements. The system reaches 290 mK without incident microwaves and has at least a one-day hold time, which is enough for one cycle of our experimental procedure.

\begin{figure}
 \includegraphics[width=0.8\columnwidth]{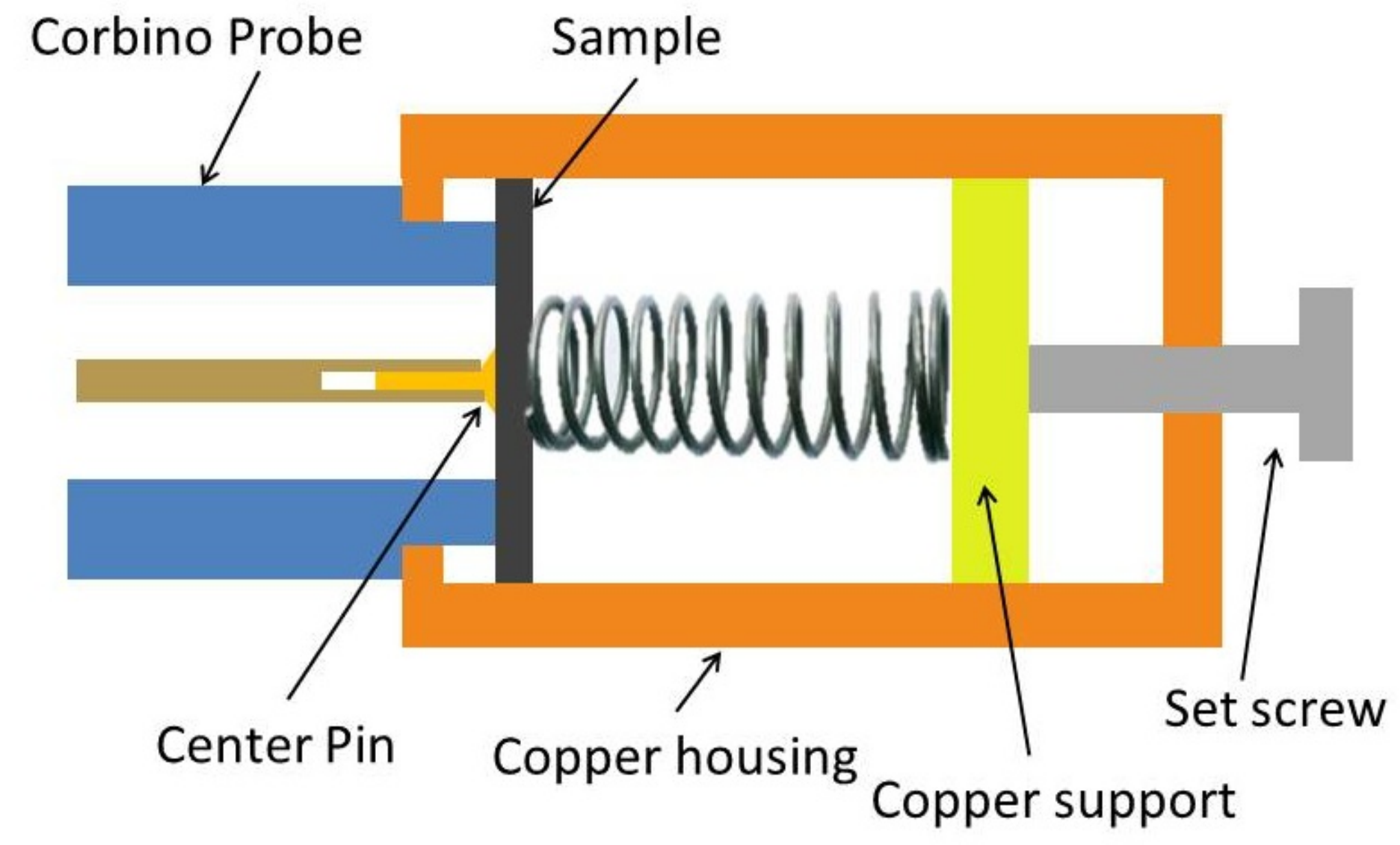}
 \caption{A schematic plot of the sample stage.   Here the Corbino Probe is the modified Rosenberger adapter.}
 \label{centerpin}
 \end{figure}

\begin{figure}
 \includegraphics[width=0.8\columnwidth]{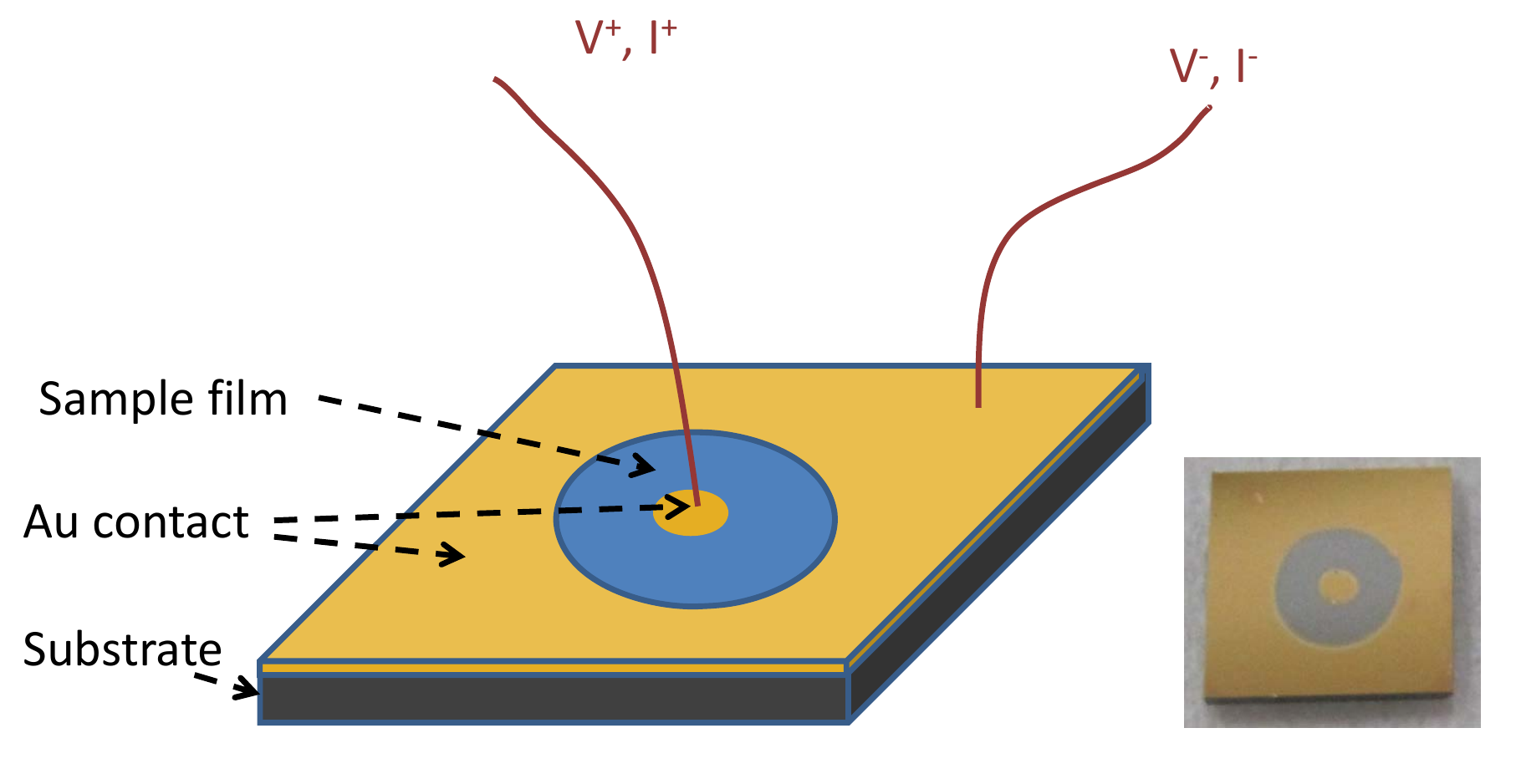}
 \caption{Gold pattern of the sample prepared for microwave measurements. A 4 probe DC measurement can be  connected to the two contacts in the Corbino geometry.   This gets rid of voltage losses in the measurement lines, but still measures an in series contact resistance. On the right we show a  picture of one of the samples we used in the experiment.}
 \label{sample4probe}
 \end{figure}

In this experiment, the complex reflection coefficients $S_{11}^m$ from the sample that terminates the transmission line at the Corbino probe (see Fig. \ref{centerpin}) are measured as a function of frequency. The Corbino probe was made from a 2.4 mm Rosenberger adapter \cite{09K121-K00S3}. We removed the threads  from one end of the adapter and carefully machined away the extra materials so that the surface of the outer conductor is flat within 0.001" deviation. Samples  are tightly pressed against the surface of the Corbino probe  to make  direct electric contact  between the outer Au pad of the film and the outer conductor of the probe (see Fig. \ref{centerpin} and Fig. \ref{sample4probe}). To bridge the gap between the center Au pad of the sample and the inner conductor of the Corbino probe (roughly about 0.005" height difference), a small brass conical shaped center pin (see Fig. \ref{centerpin}) is plugged into the inner conductor as done in reference \cite{scheffler05b}. The inner conductor of the Rosenberger adapter has 4 fingers which hold the center pin in place and also provide  necessary springy force to keep the pin in the correct configurations when a sample is attached. A 100 - 350 nm thick donut shaped gold contact was evaporated on all the samples except the open standard (see below) to reduce the contact resistant. An iron shadow mask was used and held to the sample with a magnet from the back side of the sample during gold evaporation to define the donut pattern of the sample (see Fig. \ref{sample4probe}). The inner $r_1$ and outer $r_2$ diameter of the donut shaped gold contact were 0.7 and 2.3 mm respectively.

The microwave signals come down and are reflected back along the coaxial cables. The measured reflection coefficient  $S_{11}^m$ has contributions from  losses and  phase shifts of the signals  along the coaxial cables. To obtain the real reflection coefficient $S_{11}^a$ from the sample, $S_{11}^m$ is  calibrated by measurements of three standards: open ($S_{11}^a$ = 1), load (its $S_{11}^a$ can be evaluated from its simultaneously measured DC resistance $R$ via the relation $S_{11}^a = \frac{R-Z_{0}}{R+Z_{0}}$ where $Z_{0}$ = 50 $\Omega$ is the characteristic impedance of the cable, and short ($S_{11}^a$ = -1)). The incident microwave power level was  chosen to be -27 dBm ($\sim$ 2 $\mu $W which is the lowest power that we could set the network analyzer to). This power level does not heat up the sample  at the base temperature yet gives high signal to noise.  The useful frequency range for this experiment is typically between 50 MHz to 16 GHz. Above 16 GHz, microwave reflections are dominated by the resonance in the sample holder stage or their interference  with other components in the setup. At low frequencies, usually lower than 45 MHz, microwave data appear to be contaminated by the finite contact resistance ($\sim$ 2 Ohms) of the Corbino press fit contact.

Two point DC resistance can be measured simultaneously with a lock-in amplifier by adding a bias tee \cite{Agilent 11612B Bias Network, 45 MHz to 50 GHz} to the transmission line. This bias tee sets another lower bound to the frequency range of the spectrometer.   If frequencies lower than 45 MHz are used they are passed to the lock-in which introduces substantial measurement errors to the DC measurements. Accurate determination of the resistance for the load sample is very important since the uncertainty in measuring the DC value of the load sample (either from the uncertainty in determining the contact resistance  or from inadvertently scanning the spectrometer below 45 MHz) will  propagate to errors in the conductance  of the thin films under study. The excitation currents for the lock-in amplifier during different measurements were within the range from 100 nA to 200 nA.

The interaction between the sample and the Corbino probe is  fixed by  a spring behind the sample (see Fig. \ref{centerpin}). The  overall quality of our data is improved dramatically by carefully using a caliper to set  the length of the set screw thus set the spring force to be the same for all the samples and calibration standards.

\section{Data analysis}
The actual reflection coefficient from the sample surface $S_{11}^a$ differs from the measured $S_{11}^m$ due to the effects of extraneous reflections, damping, and phase shifts in the transmission line. $S_{11}^a$ can be calculated  as:
\begin{equation}
S_{11}^a = \frac{S_{11}^m - E_D }{E_R + E_S (S_{11}^m - E_D)}.
\label{errorcoeff}
\end{equation}
\noindent Here, the frequency dependent  complex error coefficients  $E_D$, $E_S$, and $E_R$ represent the effects of directivity, attenuation, phase shifts and multiple reflections  in the transmission lines. Three reference measurements on standard samples with known reflection coefficients  are needed to determine the three unknown error coefficients. $E_D$, $E_S$, and $E_R$ at each temperature and frequency  can be determined by  solving equation (\ref{errorcoeff}) using the known reflection coefficients for the three calibration standards. $S_{11}^a$ for the sample can then be obtained via equation \ref{errorcoeff} with the extracted error coefficients. After the actual reflection coefficient has been determined, to obtain the sample sheet impedance $Z_S^{eff}$, in principle the standard equation
\begin{equation}
Z_S^{eff}= g \frac{1+S_{11}^a}{1-S_{11}^a}Z_{0}
\label{ReflectionEq}
\end{equation}
\noindent may be used.  Here $Z_{0}$ = 50 $\Omega$ is the characteristic impedance of the cable and $g = 2\pi/\ln(r_{2}/r_{1})$  is  geometric factor where $r_{2}$ and $r_{1}$ are the outer and inner diameter of the donut shaped sample (see Fig. \ref{sample4probe}). However, this $Z_{S}^{eff}$ will be the impedance of the film under study only when the substrate contribution is negligible. For a thin film where the sample thickness is much smaller than the skin depth and under the assumption that only TEM waves propagate in the transmission lines, the effective impedance for a thin film of impedance $Z_{S}$ backed by a substrate with characteristic impedance $Z_S^{Sub}$  \cite{booth96thesis} is
\begin{equation}
Z^{eff}_S = \frac{Z_S }{1+\frac{Z_S}{Z^{Sub}_S}}
\label{subcorr}
\end{equation}
 \noindent where $Z_S^{Sub}$ is the effective substrate impedance from everything that lies behind the film. For a sample that has $Z_S \ll Z^{Sub}_S$, $\frac{Z_S}{Z^{Sub}_S} \sim 0$ and equation (\ref{subcorr}) reduces to $Z^{eff}_S \cong Z_S $. But for samples that have a sheet resistance comparable to the Si substrate, the substrate impedance must be taken into account \cite{liu11a}. Therefore, in order to obtain the real response of the InO$_x$  film, it is necessary to extract $Z^{Sub}_S$. To isolate the impedance of the sample under study,  we assume that the Hagen-Rubens limit holds for InO$_x$ in the normal state since the measurement frequency is in the microwave range and it is far below the characteristic scattering frequency of most metals (in the range of many THz) \cite{liu11a}. This implies that for such a thin film in the normal state $Z_S$ should be purely  real and independent of frequency  and equal to the DC resistance. The substrate contribution $Z^{Sub}_S$ is then extracted from the calibrated InO$_x$ data at 5.6 K. With a reasonable  assumption that $Z^{Sub}_S$ of the insulating substrate is temperature independent at low temperatures, the intrinsic response of the film $Z_S$ at any other temperatures can be calculated. The complex sheet conductance $G \equiv \sigma d $ is related to sheet impedance as $G = 1/Z_S^*$ in the thin film limit \cite{liu13thesis}.

 One challenge of this calibration procedure is the repeatability of each measurement as the three error coefficients are very temperature dependent. The temperature profile along the cryostat has to be the same for the microwave measurements on three standards and the sample under study. To this end, we established a particular cool-down procedure.   Liquid nitrogen were introduced into the bath first and temperatures were allowed to equilibrate at 77 K for over 12 hours. After the initial transfer of liquid helium, the temperature of the sample stage is kept below 4 K for at least 24 hours. We adopt this receipt because we found that it took time for the superconducting cable to equilibrate.  A  slow and repeatable scan was performed for each sample from the base temperature up to 10 K in  9 hours \cite{liu13thesis}.

\section{Calibrations}
As mentioned above, three standards are needed to determine the three unknown error coefficients if one wants to know the actual response from the sample. A blank high resistivity Si substrate was used as an open standard. A 20 nm NiCr film evaporated on Si substrate was used as a load standard. NiCr thin films have a very high scattering rate, so one  can assume that the impedance of the NiCr standard is flat in our accessible frequency range.  The impedance of the NiCr films was about 42 $\Omega$, which is very closely matched to the coaxial line impedance making these excellent load standards with known reflection coefficients using the expression  $S_{11}^a = \frac{R-Z_{0}}{R+Z_{0}}$.  A 20 nm superconducting Nb film ($T_c \sim 6$ K  that can be quantified from its co-measured DC resistance) sputtered on a Si substrate was used as a short standard for zero field  measurements \cite{liu11a}. A Nb film above $T_c$ is not a good short standard due to its substantial resistance.  We found that using a superconducting Nb film on a Si substrate as a perfect short yields  more reliable results than using bulk copper  for calibrating superconducting films. This is likely the case because copper is less reflective than most bulk superconductors. Small calibration errors from the imperfection in the short standards result in a small error in the phase of the calibrated conductivity  for highly conductive samples giving a small negative contribution to the conductivity at high frequencies for some samples. However, these kinds of effects are not important for the vast majority of  samples that have dissipation  such as superconductors in the fluctuation regime, thin superconducting films in magnetic field, and graphene. Copper is a fairly good short in those cases.

For measurements in finite magnetic field, Nb films are not  perfect shorts. Therefore bulk copper with a thick Au film on top is used as  a short standard. For our measurements on superconducting InO$x$ films at finite fields \cite{liu13a}, the choice of short standards does not affect our results since superconductors become very dissipative with applied magnetic fields at finite frequency.

Short  only calibration as discussed in reference \cite{Scheffler04a} may not be possible in our setup since the three error terms have very strong temperature dependence especially after the inclusion of the superconducting cable. Different choices of open  affect the high limit of the cutoff for the usable frequency ranges. This was  discussed at length in reference \cite{Scheffler04a}. A  test of different calibration standards in our setup showed that the choice of glass, ceramic or Si substrate gave the same conclusion with regards to the experimental data \cite{liu13thesis}.

\subsection{Room temperature calibrations}

The  contact between the sample and the Corbino probe, as discussed in the previous section, is defined by the two springy forces provided by the inner conductor of the probe and the spring behind the sample. To ensure the same reference plane for all the samples,  each configuration should have the same spring force. Spring tension, which is controlled by a set screw, determines how hard the sample is pushed against the probe. Different  tensions may lead to different positions of the center pin, thus changing how the microwave radiation interacts with the sample. The main purpose of keeping  the same length of the spring beneath the sample is to maintain the same spring force for all  configurations. This approach  also ensures that the center pin will be pushed into the inner conductor at the same depth each time. We picked a standard for the length of the set screw outside of the sample stage when the sample and the Corbino probe have the best contact, which is usually defined as the point where the minimum DC resistance of the NiCr film reading is obtained.  This simultaneously   sets the force on  the spring.  The length of the set screw changes accordingly for samples with different thickness. Although the  plane of the sample surface might change once the IVC is in high vacuum  and  low temperature environments, all the reference planes should  be the same for all the samples if we have the same starting reference plane and same procedure to pump and cool down the system.

\begin{figure}
\includegraphics[width=0.8\columnwidth]{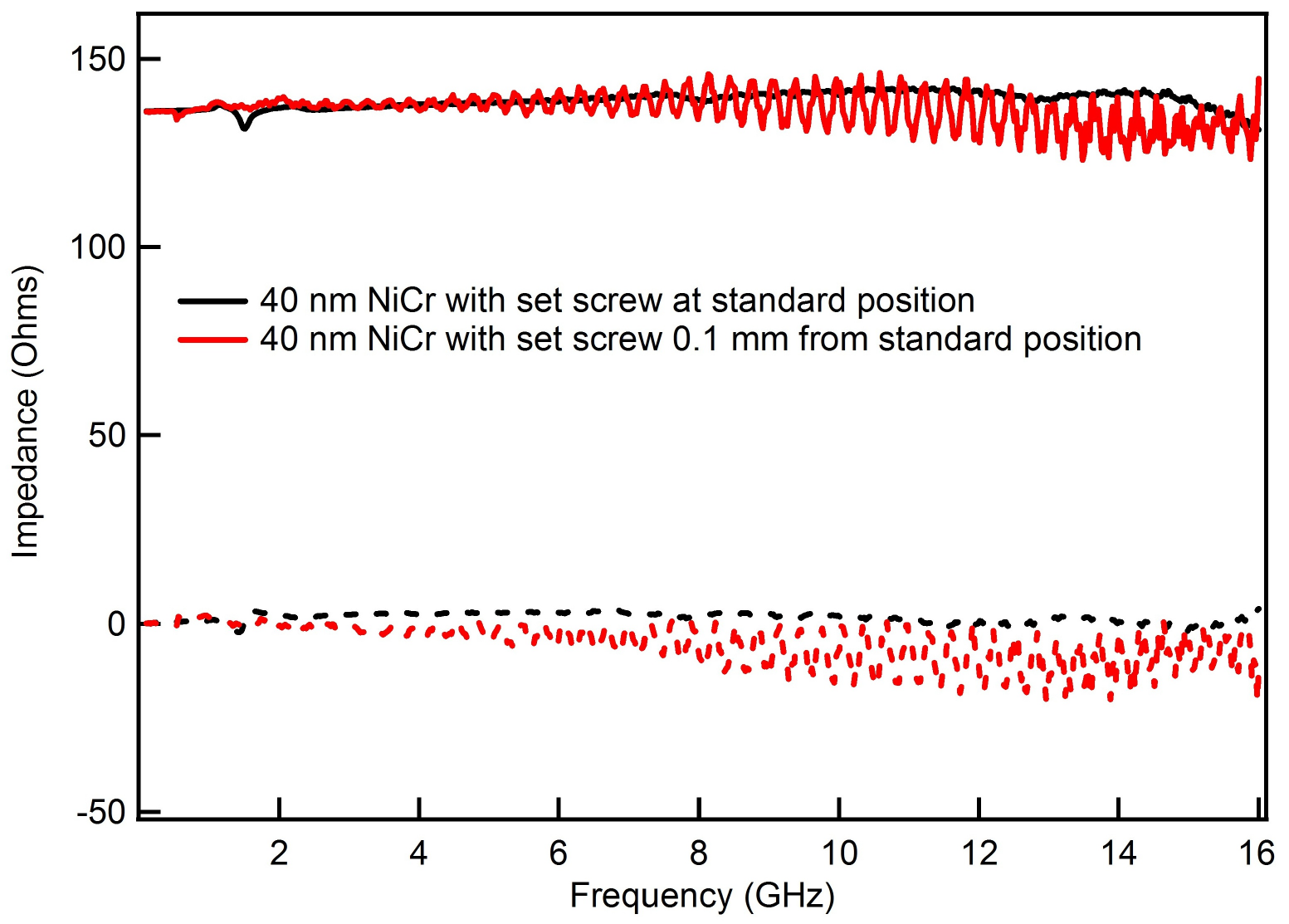}
\caption[Effects of the position of the set screw]{Real (solid lines) and imaginary (dashed lines) impedance of a 40 nm NiCr film  on a Si substrate as a function of frequency. Black curves are the frequency dependence of the calibrated real and imaginary impedance with the set screw in the standard position maintaining the same spring force for all the calibration standards and the sample under study. For the red lines, the set screw is in the right position for all the three calibration standards, but it is about 0.1 mm off from the standard position for  the 40 nm NiCr film.}
\label{comparisonofrefereceplane}
\end{figure}

As discussed above applying the same force in the spring for all the samples is  important. A large difference in the reference plane would give us different results of the sample under study.  A small difference  would give some wiggles as a function of frequency in the calibrated data as observed in Fig. \ref{comparisonofrefereceplane}. In this graph, a 40 nm NiCr film on a Si substrate was calibrated by a 20 nm NiCr film, bulk copper, and a blank Si substrate. The difference in experimental setup for the two sets of data is that the set screw was just about 0.1 mm away from its standard position for the red curves. The absolute magnitude and shape of the two sets of data are overall very similar. However, the data in red  have small oscillations in frequency in both real and imaginary impedance and this is caused by the small deviation of the reference plane when we set the set screw differently for the 40 nm NiCr film for that test run.

\begin{figure}
\includegraphics[width=\columnwidth]{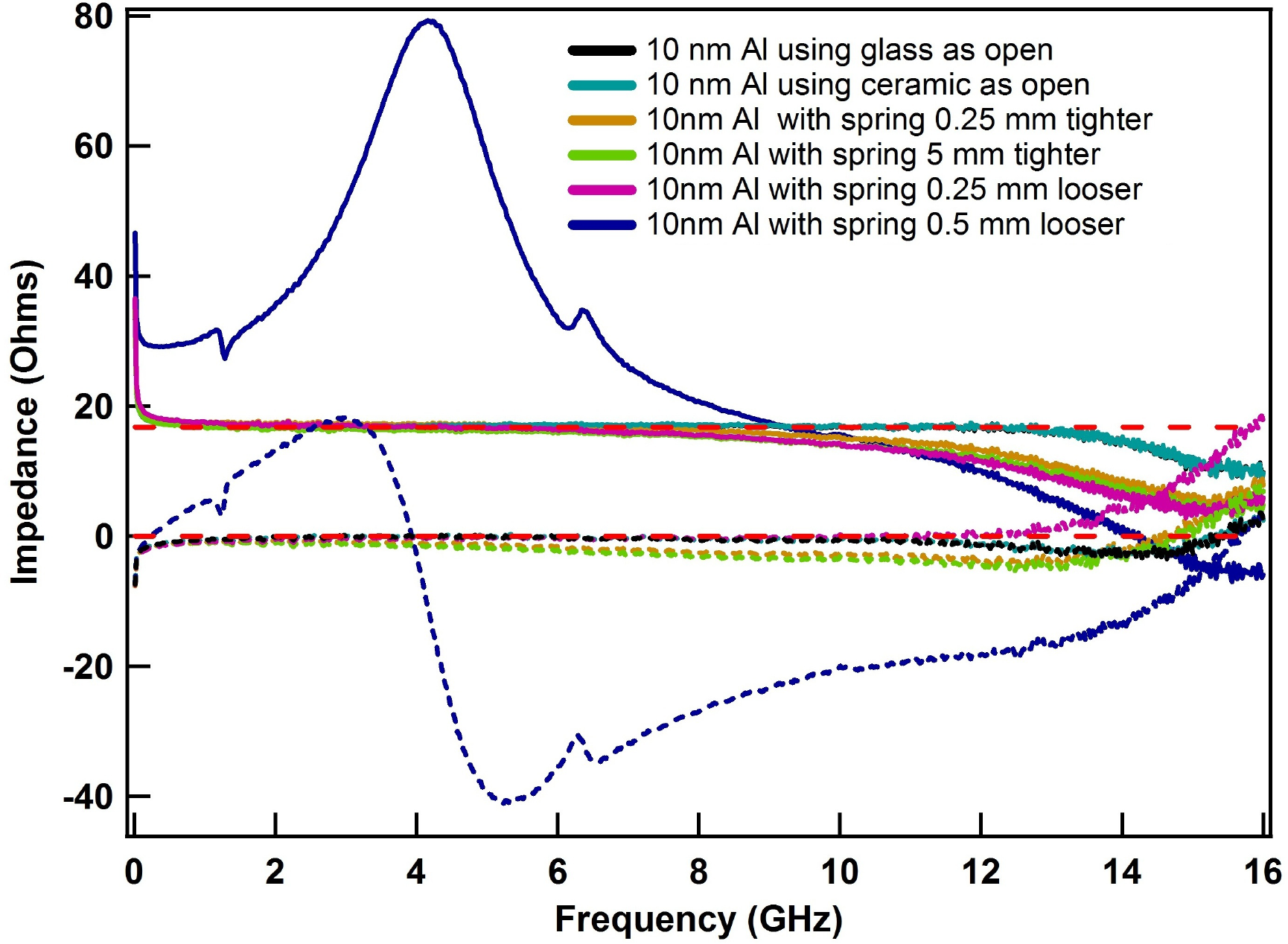}
\caption[Impedance of a 10 nm  Al  film  with different spring configurations]{Real (solid lines) and imaginary (dashed lines) impedance of a 10 nm  Al  film at room temperature with different spring configurations. The dashed lines show the expected frequency independent behavior of the impedance of the Al film at room temperature in this frequency range.}
\label{AlroomT}
\end{figure}

Fig. \ref{AlroomT} shows results for a more systematic study of the position of the set screw. A 10 nm Al film on a Si substrate was calibrated by a 20 nm NiCr film, bulk copper, and ceramic or glass. Red dashed lines are guide to the eye of a frequency independent impedance.    A standard position for the set screw was set from when the resistance  of a 20 nm NiCr film was the minimum when we change the length of the set screw. The force of the spring was adjusted to be the same by the  standard position for the three calibration standards. For the 10 nm Al film, we tried different spring configurations as described by the color legend in Fig. \ref{AlroomT}.  For  different forces in the spring, the calibrated impedance starts to slightly acquire frequency dependence,  especially at higher frequency. Here we did not distinguish which  open standard  we used  since both ceramic and glass yield the same calibrated data as demonstrated in the plot. A loose spring may result in unreliable data as shown by the dark blue curve. In that case, the sample and the probe may not even have good contact. The  drop of the real impedance at very low frequencies is the effect of    the contact resistance.

\subsection{Low temperature calibrations}
Low temperature calibrations are more complicated than room temperature calibration procedures, especially  with the superconducting cable. One experimental challenge when we were characterizing the system was the repeatability of the measurement of each sample since all the error coefficients have strong temperature dependence. To correctly remove the cable contributions, the temperature profile along the transmission line has to be the same when performing the three calibration measurements and the sample measurement. For that reason, the repeatability of the cooling down procedures for the three calibration standards and each measurement of the sample is essential. In this section, we mainly discuss the characterization of the spectrometer with the superconducting cable and demonstrate  the great repeatability of each measurement in our setup.
\subsection{Effects of the superconducting cable}
To  characterize the cables' response, we analyzed the sample's reflection coefficient's amplitude and phase separately as $S_{11}^m \equiv |S_{11}| e^{i\psi}$ ($\psi$ is in radian for the following graphs) as a function of temperature for an open  standard which should have little low temperature dependence. Therefore, the changes in $S_{11}^m$ as we scan the temperature are mainly the contributions from the cables to the reflected signals.

By characterizing $|S_{11}| e^{i\psi}$ as a function of temperature and time, we found that the superconducting cable requires a long time to reach its thermal equilibrium. We made  the cable about 10 cm long and carefully thermally anchored all the connections as detailed  in the experiment overview section above.  We strictly followed the experiment procedures \cite{liu13thesis} for each measurement cycle.   We can control the temperature repeatedly and reliably from 300 mK up to 10 K. Microwave data at zero field were taken 3 times per  cycle: first warming up (from base temperature to about 4 K), cooling down (from 20 K to 2 K) and a second warming up (from base temperature to around 10 K).   To reproduce the same temperature profile, we wait the same amount of time between each measurement for every sample.

The two warming up scans have  a difference in  $|S_{11}|$  that can be close to 2 \%, especially at low temperatures \cite{liu13thesis}. We believe this is caused by the difference in the helium volume in the dewar. The maximum difference in $|S_{11}|$ is about 2 \% for the change in the helium level between the two warming up  scans.  The difference in $|S_{11}|$  between the cooling down and second warming up  is 1 \% or less over  the whole temperature range. Although there is some difference in   $S_{11}^m$ in first warming up, cooling down and second warming up scans, the  error  due to irreproducibility between calibration scans and samples scans is less than 1 \% in $S_{11}^m$.

\begin{figure}
\includegraphics[width= 0.8\columnwidth]{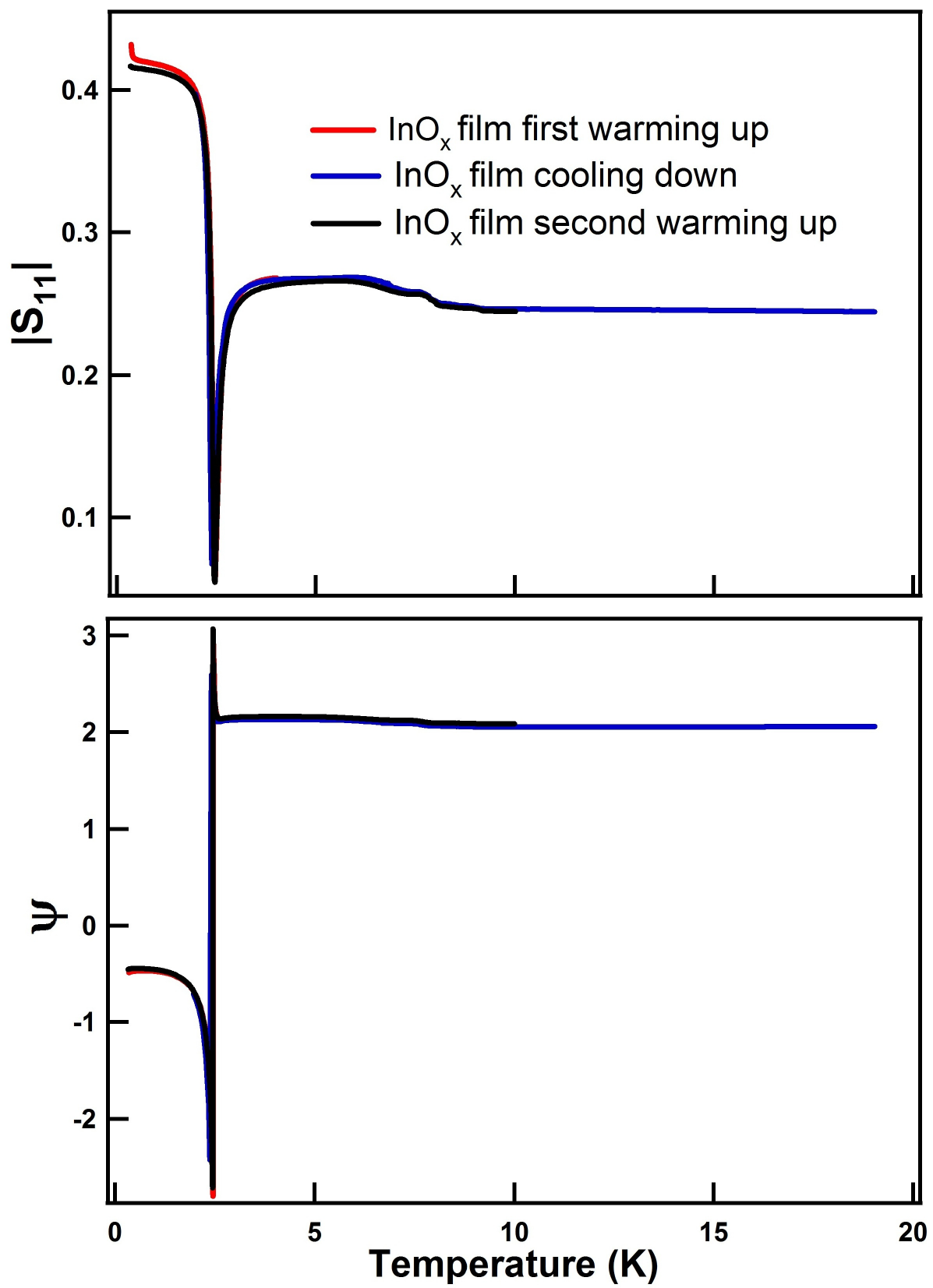}
\caption{Magnitude and phase of $S_{11}^m$ of an InO$_x$ film as a function of temperature at 7 GHz before calibration. Different runs are indicated by the color legend.}
\label{testofshortSCcableInOx}
\end{figure}

The superconducting transition in InO$_x$ film \cite{liu11a} has a very sharp feature in the raw $S_{11}^m$  data for its finite temperature superconducting transition (see Fig. \ref{testofshortSCcableInOx}).  The change in the reflection signal in the  InO$_x$ film is much larger  than the changes in the three standards over the same temperature range. We  can see the changes in the raw $S_{11}^m$ due to the existence of the transition in the  superconducting cable  around 8 $-$ 9 K, but this change is very small compared to the overall signal change in the InO$_x$ film. So error introduced by the possible difference in the helium level in the dewar should be  negligible in the final analyzed data and do not affect the physical interpretation.

\subsection{Microwave measurements in a perpendicular magnetic field}
\begin{figure}
\includegraphics[width= \columnwidth]{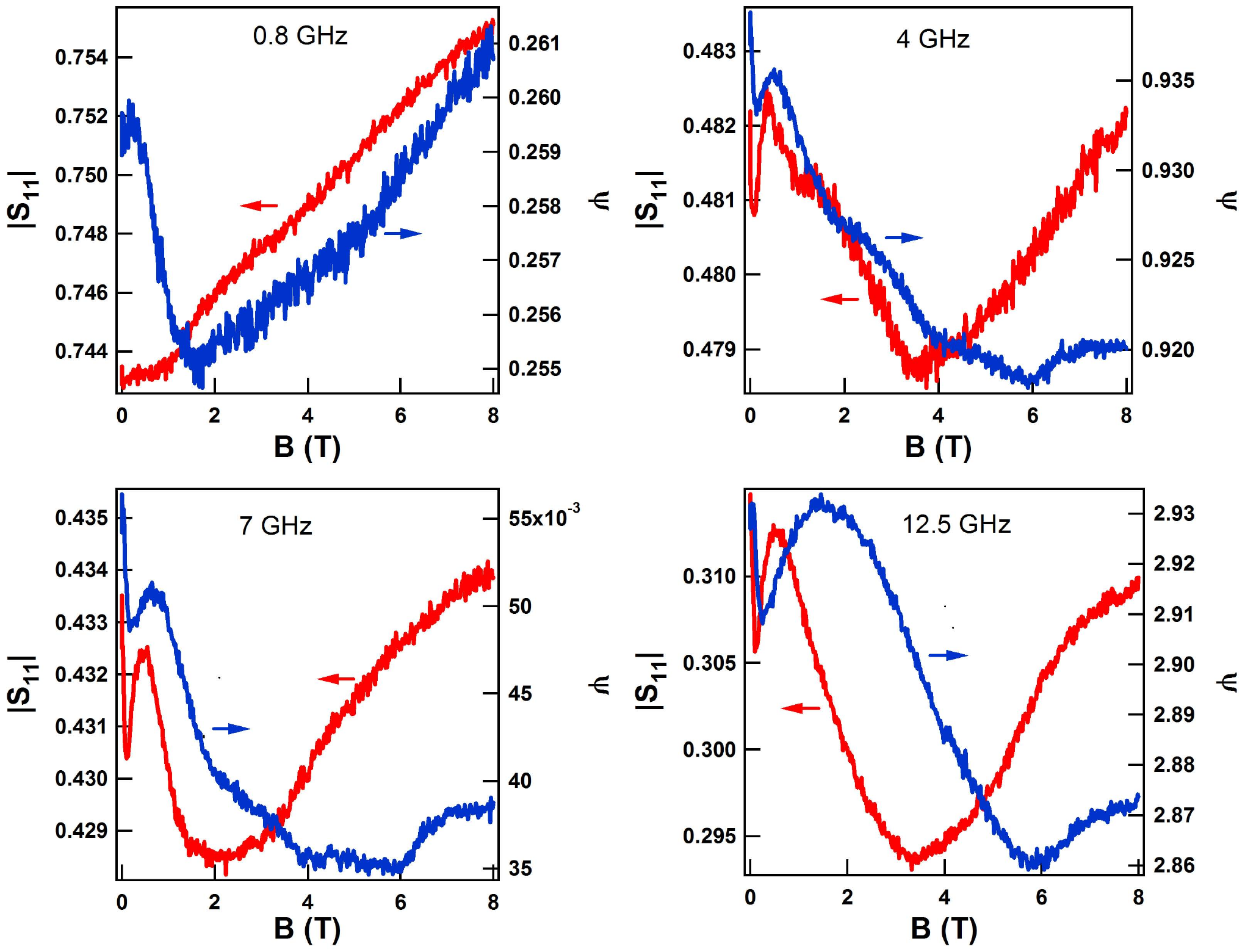}
\caption{Magnitude and phase of $S_{11}^m$ of a Si standard as a function of field at 300 mK for 0.8, 4, 7, and 12.5 GHz. Different frequency values are indicated in the legend for each plot. In all the plots, data in red are the magnitude of $S_{11}^m$ and data in blue are the phase of $S_{11}^m$ (in radians) for that particular frequency.}
\label{cableinBfield}
\end{figure}

\begin{figure}
\includegraphics[width=\columnwidth]{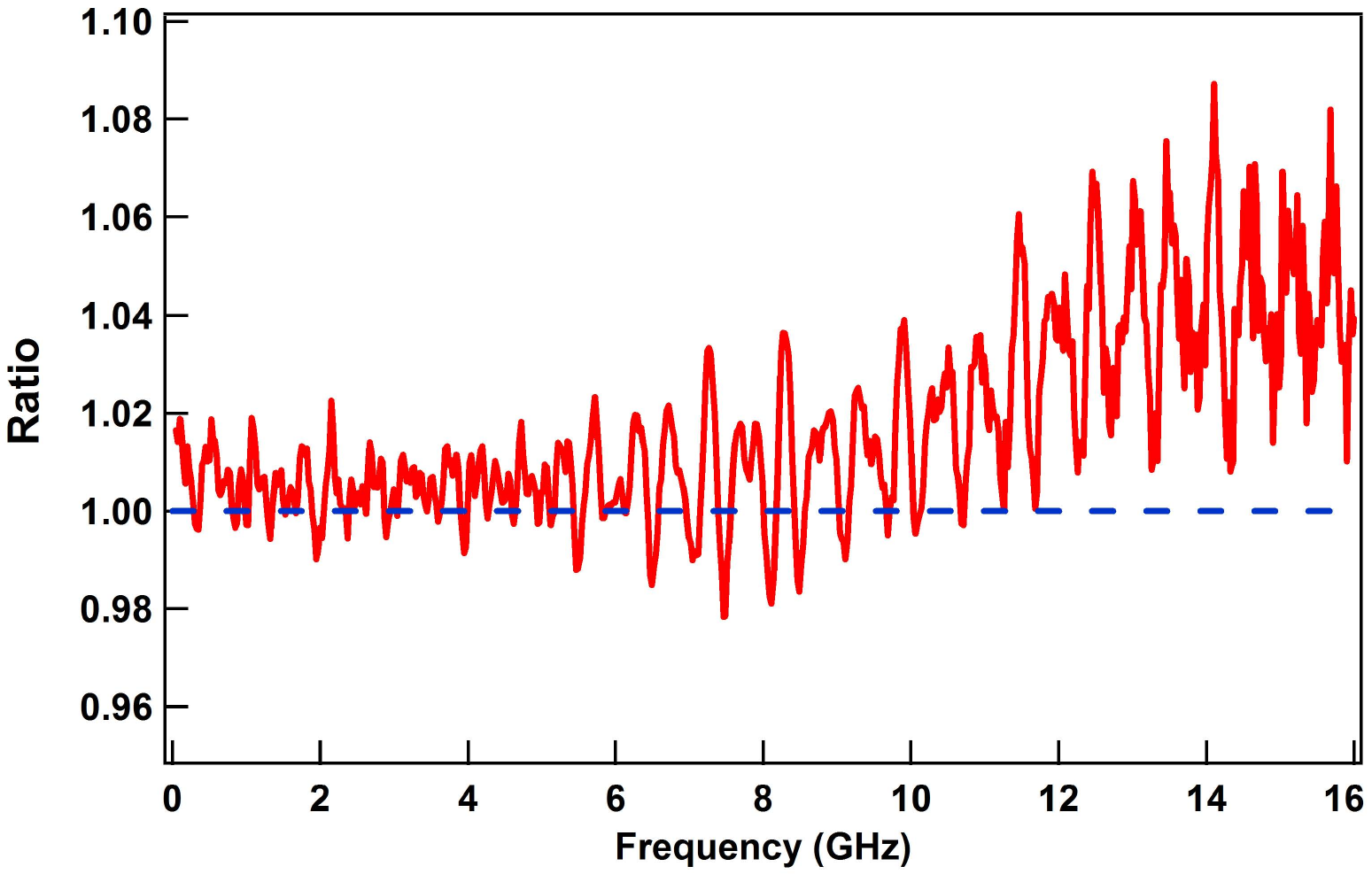}
\caption{Ratio of real conductance of InO$_x$ measured at 3.5 Tesla but calibrated by standards measured at 2 Tesla and 3.5 Tesla. Red curves are the ratios at 0.33 K. The blue dashed  line marks the expected ratio when the calibrated data from both calibration sets are the same.}
\label{calibrateddatabydiffB}
\end{figure}

For calibration purposes, we need to find out the field dependence of the coaxial cables, especially because the superconducting cable in the system might be affected by the applying magnetic field. One can in principle, examine this dependence by looking at $S_{11}^m$ of a Si standard for a magnetic field scan at fixed temperatures. Fig. \ref{cableinBfield} shows the magnitude and phase of $S_{11}^m$ of a Si standard as a function of field at 300 mK for 4 frequencies. Since the Si standard  should have little  temperature and magnetic field dependence,  the changes in $S_{11}^m$ showed in Fig. \ref{cableinBfield} for frequencies at 0.8, 4, 7, and 12.5 GHz are mainly the contributions from the cables to the reflected signals as we sweep field. Both the magnitude and phase of $S_{11}^m$   are not linear in field and they show minimums at
different fields for different frequencies. However, the overall changes in the response of the cables are still small compared with the change in the InO$_x$ signal  showed in Fig. \ref{testofshortSCcableInOx}. Fig. \ref{calibrateddatabydiffB} plots the ratios of the real conductance of InO$_x$ measured at 3.5 Tesla but calibrated by measurements of standards at 2 Tesla and 3.5 Tesla. As one can see an interpolation in field can still yield reasonable  calibrated data as long as the sample under study has a very large change in  signal at different magnetic fields. This calibration  validates the calibrated InO$_x$ data at 4 Tesla using an effective calibration interpolated from 3.5 Tesla and 5 Tesla calibration standards in reference \cite{liu13a} due to a missing set of calibration curves. These results show that it is not necessary to calibrate at each field and one can in principle calibrate sample data at a magnetic field using calibrations that  were taken at an adjacent  field value.
\section{Conclusion}
We have set up a Corbino broadband spectrometer that can obtain reliable data down to 300  mK and up to 8 Tesla in the frequency range from 50 MHz and 16 GHz. We demonstrate a way of heat sinking the inner conductor of the transmission line by including a superconducting cable. This enables us to stabilize the base temperature of the setup at 300 mK.  However, the inclusion of this superconducting cable  complicates the calibration procedure by introducing additional reflections and possible magnetic fields dependent effects.  This is overcome by strictly following a standard procedure for measurements of each sample.   Moreover, we show the importance of maintaining the same reference plane for measurements of all the standards and each sample. We are able to reproduce the temperature profile for each measurement and keep the error induced by the intrinsic complicated calibration process to be within 2\%. Reliable data have been obtained for disordered superconductors, graphene and 2D quantum phase transition in a superconductor at 300 mK \cite{liu11a,liu11b,liu13a}. We believe that this experimental apparatus has great potential for the investigation of other complex quantum states of matter at low temperatures.
\section{Acknowledgements}
The InO$_x$ sample was kindly provided by G. Sambandamurthy. We thank Marc Scheffler, Martin Dressel, Lloyd Engel and R. Valdes Aguilar for helpful discussions about the experimental setup. The research at JHU were supported by NSF DMR-0847652.


\begin{thebibliography}{30}%
\makeatletter
\providecommand \@ifxundefined [1]{%
 \@ifx{#1\undefined}
}%
\providecommand \@ifnum [1]{%
 \ifnum #1\expandafter \@firstoftwo
 \else \expandafter \@secondoftwo
 \fi
}%
\providecommand \@ifx [1]{%
 \ifx #1\expandafter \@firstoftwo
 \else \expandafter \@secondoftwo
 \fi
}%
\providecommand \natexlab [1]{#1}%
\providecommand \enquote  [1]{``#1''}%
\providecommand \bibnamefont  [1]{#1}%
\providecommand \bibfnamefont [1]{#1}%
\providecommand \citenamefont [1]{#1}%
\providecommand \href@noop [0]{\@secondoftwo}%
\providecommand \href [0]{\begingroup \@sanitize@url \@href}%
\providecommand \@href[1]{\@@startlink{#1}\@@href}%
\providecommand \@@href[1]{\endgroup#1\@@endlink}%
\providecommand \@sanitize@url [0]{\catcode `\\12\catcode `\$12\catcode
  `\&12\catcode `\#12\catcode `\^12\catcode `\_12\catcode `\%12\relax}%
\providecommand \@@startlink[1]{}%
\providecommand \@@endlink[0]{}%
\providecommand \url  [0]{\begingroup\@sanitize@url \@url }%
\providecommand \@url [1]{\endgroup\@href {#1}{\urlprefix }}%
\providecommand \urlprefix  [0]{URL }%
\providecommand \Eprint [0]{\href }%
\providecommand \doibase [0]{http://dx.doi.org/}%
\providecommand \selectlanguage [0]{\@gobble}%
\providecommand \bibinfo  [0]{\@secondoftwo}%
\providecommand \bibfield  [0]{\@secondoftwo}%
\providecommand \translation [1]{[#1]}%
\providecommand \BibitemOpen [0]{}%
\providecommand \bibitemStop [0]{}%
\providecommand \bibitemNoStop [0]{.\EOS\space}%
\providecommand \EOS [0]{\spacefactor3000\relax}%
\providecommand \BibitemShut  [1]{\csname bibitem#1\endcsname}%
\let\auto@bib@innerbib\@empty
\bibitem [{\citenamefont {Booth}, \citenamefont {Wu},\ and\ \citenamefont
  {Anlage}(1994)}]{booth94a}%
  \BibitemOpen
  \bibfield  {author} {\bibinfo {author} {\bibfnamefont {J.}~\bibnamefont
  {Booth}}, \bibinfo {author} {\bibfnamefont {D.}~\bibnamefont {Wu}}, \ and\
  \bibinfo {author} {\bibfnamefont {S.}~\bibnamefont {Anlage}},\ }\href@noop {}
  {\bibfield  {journal} {\bibinfo  {journal} {Review of Scientific
  Instruments}\ }\textbf {\bibinfo {volume} {65}},\ \bibinfo {pages} {2082}
  (\bibinfo {year} {1994})}\BibitemShut {NoStop}%
\bibitem [{\citenamefont {Wu}, \citenamefont {Booth},\ and\ \citenamefont
  {Anlage}(1995)}]{wu95}%
  \BibitemOpen
  \bibfield  {author} {\bibinfo {author} {\bibfnamefont {D.}~\bibnamefont
  {Wu}}, \bibinfo {author} {\bibfnamefont {J.}~\bibnamefont {Booth}}, \ and\
  \bibinfo {author} {\bibfnamefont {S.}~\bibnamefont {Anlage}},\ }\href@noop {}
  {\bibfield  {journal} {\bibinfo  {journal} {Physical Review Letters}\
  }\textbf {\bibinfo {volume} {75}},\ \bibinfo {pages} {525} (\bibinfo {year}
  {1995})}\BibitemShut {NoStop}%
\bibitem [{\citenamefont {Anlage}\ \emph {et~al.}(1996)\citenamefont {Anlage},
  \citenamefont {Mao}, \citenamefont {Booth}, \citenamefont {Wu},\ and\
  \citenamefont {Peng}}]{anlage96}%
  \BibitemOpen
  \bibfield  {author} {\bibinfo {author} {\bibfnamefont {S.}~\bibnamefont
  {Anlage}}, \bibinfo {author} {\bibfnamefont {J.}~\bibnamefont {Mao}},
  \bibinfo {author} {\bibfnamefont {J.}~\bibnamefont {Booth}}, \bibinfo
  {author} {\bibfnamefont {D.}~\bibnamefont {Wu}}, \ and\ \bibinfo {author}
  {\bibfnamefont {J.}~\bibnamefont {Peng}},\ }\href@noop {} {\bibfield
  {journal} {\bibinfo  {journal} {Physical Review B}\ }\textbf {\bibinfo
  {volume} {53}},\ \bibinfo {pages} {2792} (\bibinfo {year}
  {1996})}\BibitemShut {NoStop}%
\bibitem [{\citenamefont {Booth}\ \emph {et~al.}(1996)\citenamefont {Booth},
  \citenamefont {Wu}, \citenamefont {Qadri}, \citenamefont {Skelton},
  \citenamefont {Osofsky}, \citenamefont {Piqu\'e},\ and\ \citenamefont
  {Anlage}}]{BoothPRL96a}%
  \BibitemOpen
  \bibfield  {author} {\bibinfo {author} {\bibfnamefont {J.~C.}\ \bibnamefont
  {Booth}}, \bibinfo {author} {\bibfnamefont {D.~H.}\ \bibnamefont {Wu}},
  \bibinfo {author} {\bibfnamefont {S.~B.}\ \bibnamefont {Qadri}}, \bibinfo
  {author} {\bibfnamefont {E.~F.}\ \bibnamefont {Skelton}}, \bibinfo {author}
  {\bibfnamefont {M.~S.}\ \bibnamefont {Osofsky}}, \bibinfo {author}
  {\bibfnamefont {A.}~\bibnamefont {Piqu\'e}}, \ and\ \bibinfo {author}
  {\bibfnamefont {S.~M.}\ \bibnamefont {Anlage}},\ }\href {\doibase
  10.1103/PhysRevLett.77.4438} {\bibfield  {journal} {\bibinfo  {journal}
  {Physical Review Letters}\ }\textbf {\bibinfo {volume} {77}},\ \bibinfo
  {pages} {4438} (\bibinfo {year} {1996})}\BibitemShut {NoStop}%
\bibitem [{\citenamefont {Xu}\ \emph {et~al.}(2009)\citenamefont {Xu},
  \citenamefont {Li}, \citenamefont {Anlage}, \citenamefont {Lobb},
  \citenamefont {Sullivan}, \citenamefont {Segawa},\ and\ \citenamefont
  {Ando}}]{xu09}%
  \BibitemOpen
  \bibfield  {author} {\bibinfo {author} {\bibfnamefont {H.}~\bibnamefont
  {Xu}}, \bibinfo {author} {\bibfnamefont {S.}~\bibnamefont {Li}}, \bibinfo
  {author} {\bibfnamefont {S.}~\bibnamefont {Anlage}}, \bibinfo {author}
  {\bibfnamefont {C.}~\bibnamefont {Lobb}}, \bibinfo {author} {\bibfnamefont
  {M.}~\bibnamefont {Sullivan}}, \bibinfo {author} {\bibfnamefont
  {K.}~\bibnamefont {Segawa}}, \ and\ \bibinfo {author} {\bibfnamefont
  {Y.}~\bibnamefont {Ando}},\ }\href@noop {} {\bibfield  {journal} {\bibinfo
  {journal} {Physical Review B}\ }\textbf {\bibinfo {volume} {80}},\ \bibinfo
  {pages} {104518} (\bibinfo {year} {2009})}\BibitemShut {NoStop}%
\bibitem [{\citenamefont {Schwartz}, \citenamefont {Scheffler},\ and\
  \citenamefont {Anlage}(2000)}]{schwartz00}%
  \BibitemOpen
  \bibfield  {author} {\bibinfo {author} {\bibfnamefont {A.}~\bibnamefont
  {Schwartz}}, \bibinfo {author} {\bibfnamefont {M.}~\bibnamefont {Scheffler}},
  \ and\ \bibinfo {author} {\bibfnamefont {S.}~\bibnamefont {Anlage}},\
  }\href@noop {} {\bibfield  {journal} {\bibinfo  {journal} {Physical Review
  B}\ }\textbf {\bibinfo {volume} {61}},\ \bibinfo {pages} {870} (\bibinfo
  {year} {2000})}\BibitemShut {NoStop}%
\bibitem [{\citenamefont {Lee}\ and\ \citenamefont
  {Stutzmann}(2001)}]{LeePRL01b}%
  \BibitemOpen
  \bibfield  {author} {\bibinfo {author} {\bibfnamefont {M.}~\bibnamefont
  {Lee}}\ and\ \bibinfo {author} {\bibfnamefont {M.~L.}\ \bibnamefont
  {Stutzmann}},\ }\href {\doibase 10.1103/PhysRevLett.87.056402} {\bibfield
  {journal} {\bibinfo  {journal} {Physical Review Letters}\ }\textbf {\bibinfo
  {volume} {87}},\ \bibinfo {pages} {056402} (\bibinfo {year}
  {2001})}\BibitemShut {NoStop}%
\bibitem [{\citenamefont {Martens}, \citenamefont {Reedijk},\ and\
  \citenamefont {Brom}(2000)}]{martens00}%
  \BibitemOpen
  \bibfield  {author} {\bibinfo {author} {\bibfnamefont {H.}~\bibnamefont
  {Martens}}, \bibinfo {author} {\bibfnamefont {J.}~\bibnamefont {Reedijk}}, \
  and\ \bibinfo {author} {\bibfnamefont {H.}~\bibnamefont {Brom}},\ }\href@noop
  {} {\bibfield  {journal} {\bibinfo  {journal} {Review of Scientific
  Instruments}\ }\textbf {\bibinfo {volume} {71}},\ \bibinfo {pages} {473}
  (\bibinfo {year} {2000})}\BibitemShut {NoStop}%
\bibitem [{\citenamefont {Scheffler}\ and\ \citenamefont
  {Dressel}(2005)}]{scheffler05b}%
  \BibitemOpen
  \bibfield  {author} {\bibinfo {author} {\bibfnamefont {M.}~\bibnamefont
  {Scheffler}}\ and\ \bibinfo {author} {\bibfnamefont {M.}~\bibnamefont
  {Dressel}},\ }\href@noop {} {\bibfield  {journal} {\bibinfo  {journal}
  {Review of Scientific Instruments}\ }\textbf {\bibinfo {volume} {76}},\
  \bibinfo {pages} {074702} (\bibinfo {year} {2005})}\BibitemShut {NoStop}%
\bibitem [{\citenamefont {Scheffler}\ \emph {et~al.}(2005)\citenamefont
  {Scheffler}, \citenamefont {Dressel}, \citenamefont {Martin},\ and\
  \citenamefont {Adrian}}]{Scheffler05a}%
  \BibitemOpen
  \bibfield  {author} {\bibinfo {author} {\bibfnamefont {M.}~\bibnamefont
  {Scheffler}}, \bibinfo {author} {\bibfnamefont {M.}~\bibnamefont {Dressel}},
  \bibinfo {author} {\bibfnamefont {J.}~\bibnamefont {Martin}}, \ and\ \bibinfo
  {author} {\bibfnamefont {H.}~\bibnamefont {Adrian}},\ }\href {\doibase
  10.1038/nature04232} {\bibfield  {journal} {\bibinfo  {journal} {Nature
  (London)}\ }\textbf {\bibinfo {volume} {438}},\ \bibinfo {pages} {1135}
  (\bibinfo {year} {2005})}\BibitemShut {NoStop}%
\bibitem [{\citenamefont {Steinberg}, \citenamefont {Scheffler},\ and\
  \citenamefont {Dressel}(2008)}]{Scheffler08a}%
  \BibitemOpen
  \bibfield  {author} {\bibinfo {author} {\bibfnamefont {K.}~\bibnamefont
  {Steinberg}}, \bibinfo {author} {\bibfnamefont {M.}~\bibnamefont
  {Scheffler}}, \ and\ \bibinfo {author} {\bibfnamefont {M.}~\bibnamefont
  {Dressel}},\ }\href@noop {} {\bibfield  {journal} {\bibinfo  {journal}
  {Physical Review B}\ }\textbf {\bibinfo {volume} {77}},\ \bibinfo {pages}
  {214517} (\bibinfo {year} {2008})}\BibitemShut {NoStop}%
\bibitem [{\citenamefont {Kitano}, \citenamefont {Ohashi},\ and\ \citenamefont
  {Maeda}(2008)}]{kitano08a}%
  \BibitemOpen
  \bibfield  {author} {\bibinfo {author} {\bibfnamefont {H.}~\bibnamefont
  {Kitano}}, \bibinfo {author} {\bibfnamefont {T.}~\bibnamefont {Ohashi}}, \
  and\ \bibinfo {author} {\bibfnamefont {A.}~\bibnamefont {Maeda}},\
  }\href@noop {} {\bibfield  {journal} {\bibinfo  {journal} {Review of
  Scientific Instruments}\ }\textbf {\bibinfo {volume} {79}},\ \bibinfo {pages}
  {074701} (\bibinfo {year} {2008})}\BibitemShut {NoStop}%
\bibitem [{\citenamefont {Kitano}\ \emph {et~al.}(2007)\citenamefont {Kitano},
  \citenamefont {Ohashi}, \citenamefont {Maeda},\ and\ \citenamefont
  {Tsukada}}]{kitano07a}%
  \BibitemOpen
  \bibfield  {author} {\bibinfo {author} {\bibfnamefont {H.}~\bibnamefont
  {Kitano}}, \bibinfo {author} {\bibfnamefont {T.}~\bibnamefont {Ohashi}},
  \bibinfo {author} {\bibfnamefont {A.}~\bibnamefont {Maeda}}, \ and\ \bibinfo
  {author} {\bibfnamefont {I.}~\bibnamefont {Tsukada}},\ }\href@noop {}
  {\bibfield  {journal} {\bibinfo  {journal} {Physica C: Superconductivity}\
  }\textbf {\bibinfo {volume} {460}},\ \bibinfo {pages} {904} (\bibinfo {year}
  {2007})}\BibitemShut {NoStop}%
\bibitem [{\citenamefont {Ohashi}\ \emph {et~al.}(2009)\citenamefont {Ohashi},
  \citenamefont {Kitano}, \citenamefont {Tsukada},\ and\ \citenamefont
  {Maeda}}]{KitanoPRB09a}%
  \BibitemOpen
  \bibfield  {author} {\bibinfo {author} {\bibfnamefont {T.}~\bibnamefont
  {Ohashi}}, \bibinfo {author} {\bibfnamefont {H.}~\bibnamefont {Kitano}},
  \bibinfo {author} {\bibfnamefont {I.}~\bibnamefont {Tsukada}}, \ and\
  \bibinfo {author} {\bibfnamefont {A.}~\bibnamefont {Maeda}},\ }\href
  {\doibase 10.1103/PhysRevB.79.184507} {\bibfield  {journal} {\bibinfo
  {journal} {Physical Review B}\ }\textbf {\bibinfo {volume} {79}},\ \bibinfo
  {pages} {184507} (\bibinfo {year} {2009})}\BibitemShut {NoStop}%
\bibitem [{\citenamefont {Ohashi}\ \emph {et~al.}(2006)\citenamefont {Ohashi},
  \citenamefont {Kitano}, \citenamefont {Maeda}, \citenamefont {Akaike},\ and\
  \citenamefont {Fujimaki}}]{ohashi06}%
  \BibitemOpen
  \bibfield  {author} {\bibinfo {author} {\bibfnamefont {T.}~\bibnamefont
  {Ohashi}}, \bibinfo {author} {\bibfnamefont {H.}~\bibnamefont {Kitano}},
  \bibinfo {author} {\bibfnamefont {A.}~\bibnamefont {Maeda}}, \bibinfo
  {author} {\bibfnamefont {H.}~\bibnamefont {Akaike}}, \ and\ \bibinfo {author}
  {\bibfnamefont {A.}~\bibnamefont {Fujimaki}},\ }\href@noop {} {\bibfield
  {journal} {\bibinfo  {journal} {Physical Review B}\ }\textbf {\bibinfo
  {volume} {73}},\ \bibinfo {pages} {174522} (\bibinfo {year}
  {2006})}\BibitemShut {NoStop}%
\bibitem [{\citenamefont {Mondal}\ \emph {et~al.}(2013)\citenamefont {Mondal},
  \citenamefont {Kamlapure}, \citenamefont {Ganguli}, \citenamefont
  {Jesudasan}, \citenamefont {Bagwe}, \citenamefont {Benfatto},\ and\
  \citenamefont {Raychaudhuri}}]{mondal13}%
  \BibitemOpen
  \bibfield  {author} {\bibinfo {author} {\bibfnamefont {M.}~\bibnamefont
  {Mondal}}, \bibinfo {author} {\bibfnamefont {A.}~\bibnamefont {Kamlapure}},
  \bibinfo {author} {\bibfnamefont {S.~C.}\ \bibnamefont {Ganguli}}, \bibinfo
  {author} {\bibfnamefont {J.}~\bibnamefont {Jesudasan}}, \bibinfo {author}
  {\bibfnamefont {V.}~\bibnamefont {Bagwe}}, \bibinfo {author} {\bibfnamefont
  {L.}~\bibnamefont {Benfatto}}, \ and\ \bibinfo {author} {\bibfnamefont
  {P.}~\bibnamefont {Raychaudhuri}},\ }\href@noop {} {\bibfield  {journal}
  {\bibinfo  {journal} {Scientific Reports}\ }\textbf {\bibinfo {volume} {3}}
  (\bibinfo {year} {2013})}\BibitemShut {NoStop}%
\bibitem [{\citenamefont {Liu}\ \emph {et~al.}(2011{\natexlab{a}})\citenamefont
  {Liu}, \citenamefont {Kim}, \citenamefont {Sambandamurthy},\ and\
  \citenamefont {Armitage}}]{liu11a}%
  \BibitemOpen
  \bibfield  {author} {\bibinfo {author} {\bibfnamefont {W.}~\bibnamefont
  {Liu}}, \bibinfo {author} {\bibfnamefont {M.}~\bibnamefont {Kim}}, \bibinfo
  {author} {\bibfnamefont {G.}~\bibnamefont {Sambandamurthy}}, \ and\ \bibinfo
  {author} {\bibfnamefont {N.~P.}\ \bibnamefont {Armitage}},\ }\href@noop {}
  {\bibfield  {journal} {\bibinfo  {journal} {Physical Review B}\ }\textbf
  {\bibinfo {volume} {84}} (\bibinfo {year} {2011}{\natexlab{a}})}\BibitemShut
  {NoStop}%
\bibitem [{\citenamefont {Liu}\ \emph {et~al.}(2013)\citenamefont {Liu},
  \citenamefont {Pan}, \citenamefont {Wen}, \citenamefont {Kim}, \citenamefont
  {Sambandamurthy},\ and\ \citenamefont {Armitage}}]{liu13a}%
  \BibitemOpen
  \bibfield  {author} {\bibinfo {author} {\bibfnamefont {W.}~\bibnamefont
  {Liu}}, \bibinfo {author} {\bibfnamefont {L.}~\bibnamefont {Pan}}, \bibinfo
  {author} {\bibfnamefont {J.}~\bibnamefont {Wen}}, \bibinfo {author}
  {\bibfnamefont {M.}~\bibnamefont {Kim}}, \bibinfo {author} {\bibfnamefont
  {G.}~\bibnamefont {Sambandamurthy}}, \ and\ \bibinfo {author} {\bibfnamefont
  {N.}~\bibnamefont {Armitage}},\ }\href@noop {} {\bibfield  {journal}
  {\bibinfo  {journal} {Physical Review Letters}\ }\textbf {\bibinfo {volume}
  {111}},\ \bibinfo {pages} {067003} (\bibinfo {year} {2013})}\BibitemShut
  {NoStop}%
\bibitem [{\citenamefont {Liu}\ \emph {et~al.}(2011{\natexlab{b}})\citenamefont
  {Liu}, \citenamefont {Aguilar}, \citenamefont {Hao}, \citenamefont {Ruoff},\
  and\ \citenamefont {Armitage}}]{liu11b}%
  \BibitemOpen
  \bibfield  {author} {\bibinfo {author} {\bibfnamefont {W.}~\bibnamefont
  {Liu}}, \bibinfo {author} {\bibfnamefont {R.}~\bibnamefont {Aguilar}},
  \bibinfo {author} {\bibfnamefont {Y.}~\bibnamefont {Hao}}, \bibinfo {author}
  {\bibfnamefont {R.}~\bibnamefont {Ruoff}}, \ and\ \bibinfo {author}
  {\bibfnamefont {N.}~\bibnamefont {Armitage}},\ }\href@noop {} {\bibfield
  {journal} {\bibinfo  {journal} {Journal of Applied Physics}\ }\textbf
  {\bibinfo {volume} {110}},\ \bibinfo {pages} {083510} (\bibinfo {year}
  {2011}{\natexlab{b}})}\BibitemShut {NoStop}%
\bibitem [{\citenamefont {Steinberg}, \citenamefont {Scheffler},\ and\
  \citenamefont {Dressel}(2012)}]{steinberg12a}%
  \BibitemOpen
  \bibfield  {author} {\bibinfo {author} {\bibfnamefont {K.}~\bibnamefont
  {Steinberg}}, \bibinfo {author} {\bibfnamefont {M.}~\bibnamefont
  {Scheffler}}, \ and\ \bibinfo {author} {\bibfnamefont {M.}~\bibnamefont
  {Dressel}},\ }\href@noop {} {\bibfield  {journal} {\bibinfo  {journal}
  {Review of Scientific Instruments}\ }\textbf {\bibinfo {volume} {83}},\
  \bibinfo {pages} {024704} (\bibinfo {year} {2012})}\BibitemShut {NoStop}%
\bibitem [{Note1()}]{Note1}%
  \BibitemOpen
  \bibinfo {note} {Agilent N5230A}\BibitemShut {NoStop}%
\bibitem [{Note2()}]{Note2}%
  \BibitemOpen
  \bibinfo {note} {Micro-Coax Company, UT-85C-TP-LL}\BibitemShut {NoStop}%
\bibitem [{Note3()}]{Note3}%
  \BibitemOpen
  \bibinfo {note} {Micro-Coax Company, UT-085-SS}\BibitemShut {NoStop}%
\bibitem [{Note4()}]{Note4}%
  \BibitemOpen
  \bibinfo {note} {Keycom Company, NbTiNbTi085A}\BibitemShut {NoStop}%
\bibitem [{Note5()}]{Note5}%
  \BibitemOpen
  \bibinfo {note} {Kawashima Manufacturing Company, KPC185FFHA}\BibitemShut
  {NoStop}%
\bibitem [{Note6()}]{Note6}%
  \BibitemOpen
  \bibinfo {note} {09K121-K00S3}\BibitemShut {NoStop}%
\bibitem [{Note7()}]{Note7}%
  \BibitemOpen
  \bibinfo {note} {Agilent 11612B Bias Network, 45 MHz to 50 GHz}\BibitemShut
  {NoStop}%
\bibitem [{\citenamefont {Booth}(1996)}]{booth96thesis}%
  \BibitemOpen
  \bibfield  {author} {\bibinfo {author} {\bibfnamefont {J.}~\bibnamefont
  {Booth}},\ }\emph {\bibinfo {title} {{Novel measurements of the frequency
  dependent microwave surface impedance of cuprate thin film
  superconductors}}},\ \href@noop {} {Ph.D. thesis},\ \bibinfo  {school}
  {University of Maryland at College Park} (\bibinfo {year} {1996})\BibitemShut
  {NoStop}%
\bibitem [{\citenamefont {Liu}(2013)}]{liu13thesis}%
  \BibitemOpen
  \bibfield  {author} {\bibinfo {author} {\bibfnamefont {W.}~\bibnamefont
  {Liu}},\ }\emph {\bibinfo {title} {{Broadband microwave measurements of two
  dimensional quantum matter}}},\ \href@noop {} {Ph.D. thesis},\ \bibinfo
  {school} {Johns Hopkins University} (\bibinfo {year} {2013})\BibitemShut
  {NoStop}%
\bibitem [{\citenamefont {Scheffler}(2004)}]{Scheffler04a}%
  \BibitemOpen
  \bibfield  {author} {\bibinfo {author} {\bibfnamefont {M.}~\bibnamefont
  {Scheffler}},\ }\emph {\bibinfo {title} {Broadband Microwave Spectroscopy on
  Correlated Electrons}},\ \href@noop {} {Ph.D. thesis},\ \bibinfo  {school}
  {University of Stuttgart} (\bibinfo {year} {2004})\BibitemShut {NoStop}%
\end{thebibliography}%
\end{document}